# Magnetic cycles and photometric variability of T Tauri stars

P.J. Armitage
*Institute of Astronomy, Madingley Road, Cambridge CB3 0HA*

**ABSTRACT**

We consider the effects of stellar magnetic cycles on disc accretion in T Tauri stars where the inner region of the disc is disrupted by the stellar field. Using time-dependent disc models, we show that order unity variations in field strength can strongly modulate the accretion flow for cycle time-scales of years to decades. Photometric variability is expected to be greatest in the ultraviolet, with smaller but still significant variations probable in the near-infrared. We suggest that variable stellar magnetic fields may contribute to the long-time-scale photometric variability of T Tauri stars. The long-term variability of T Tauri stars may therefore provide a clue to the nature of magnetic dynamo cycles in these stars.

## 1 INTRODUCTION

The spectral properties and photometric variability of pre-main-sequence stars provide valuable clues to their circumstellar environment. In the classical T Tauri stars (CTTs), prominent ultraviolet and infrared excesses are taken as indicating that these systems are actively accreting matter from surrounding discs. The simplest picture of this process, in which the disc terminates in a boundary layer at the stellar equator, must be modified in the case where the star possesses a strong magnetic field. Such models were discussed in the context of magnetized neutron stars (Ghosh & Lamb 1979), and have recently been applied to the case of T Tauri stars (Königl 1991; Cameron & Campbell 1993; Yi 1994; Ghosh 1995; Cameron, Campbell & Quaintrell 1995). The magnetic field provides a mechanism that couples the star to the inner regions of the disc, and which can disrupt the disc entirely at a few stellar radii if the field is strong enough. Inward of this radius the inflowing matter leaves the disc plane and follows the field lines to strike the star in a high-latitude accretion column. A variety of estimates suggest that the critical stellar dipole field required to disrupt a typical T Tauri disc is a few hundred gauss. The presence of ordered fields of the required strength remains unconfirmed, but would be consistent with the strong X-ray activity observed in T Tauri stars (Montmerle et al. 1992). Upper limits exist for some T Tauri stars from Zeeman effect measurements (Johnstone & Penston 1987; Basri, Marcy & Valenti 1992).

A number of observations provide support for a magnetically dominated accretion model. The coupling of the star to its disc by the field allows the star to accrete matter while rotating well below break-up, and leads to correlations between the presence or absence of discs and spin periods that are seen in surveys of T Tauri spin periods (Bouvier et al. 1993; Edwards et al. 1993). Time series photometry has been used to infer the presence of spots hotter than the stellar photosphere in several CTTs (Vrba et al. 1993; Bertout, Basri & Bouvier 1988; Simon, Vrba & Herbst 1990; Bouvier et al. 1993), consistent with the high-latitude accretion shocks in a magnetospheric model. The existence of accretion columns is also suggested by modelling of spectral lines, the profiles of which imply they are formed in an infalling rather than boundary layer geometry (Calvet & Hartmann 1992; Hartmann, Hewett & Calvet 1994). Further support is provided by the near-infrared spectral energy distributions of some CTTs, which are best fitted by models with discs truncated at a few stellar radii (Bertout et al. 1988; Edwards et al. 1993). A recent discussion of these and other observations (Hartmann 1994) concludes that magnetospheric rather than boundary layer accretion may well be the norm in T Tauri stars.

In this paper we extend the model to incorporate the effects of a variable stellar magnetic field. Our motivation is the observed variability of T Tauri stars on time-scales of years and decades (Herbig & Bell 1988). These time-scales are short compared with those for variations in stellar luminosity or disc flow at large radius, but are typical of the time-scales for magnetic cycles seen in cool stars (Baliunas 1988). The varying field modulates the accretion flow in the inner regions of the disc, producing changes in the strength of both the infrared disc emission and the spectral lines formed in the infalling accretion column. If the magnetosphere lies beyond the corotation radius (where the Keplerian disc material is stationary in a frame rotating with the star) then accretion is cut off altogether, and the star would be seen to have the spectral character of a weak-lined T Tauri star (WTT). We thus argue that magnetic cycles where the magnetosphere crosses the corotation radius are a possible mechanism for the interchange of T Tauri stars between weak-lined and classical status.

In Section 2 we describe the incorporation of stellar magnetic torques in a time-dependent disc model. In Section 3 we present model calculations for parameters we believe typical of T Tauri star-disc systems, and discuss the limits on cycle amplitude and period for which observable effects are expected. We also consider how our results depend on the viscosity in T Tauri discs, here as elsewhere poorly known. In Section 4 we discuss what might be learned from observing such cycles, and summarize our conclusions.

## 2  DESCRIPTION OF THE MODEL

### 2.1  Magnetic field configuration

We model the unperturbed (in the absence of a disc) stellar magnetic field as a dipole, aligned with the stellar rotation axis and perpendicular to the disc plane. Close to the stellar surface the field structure is doubtless complex, and for simplicity we assume that in the region of interest at a few stellar radii the dipole component is dominant. For an aligned dipole/rotation axis the problem remains axisymmetric; we thus adopt cylindrical polar coordinates $(R, \phi, z)$. Following previous authors (Ghosh & Lamb 1979) we argue that, despite the high electrical conductivity in the inner disc regions, the stellar field is able to penetrate the disc fully at all radii, as a result of local instabilities that mix the magnetospheric field with the disc plasma. Then the magnitude of the magnetic torque exerted on an annulus of the disc of width $\Delta R$ at radius $R$ is

$$T_B = B_z B_\phi R^2 \Delta R, \qquad (1)$$

where $B_z$ and $B_\phi$ are the magnetic field components evaluated at the disc surface.

We take for the $B_z$ in equation (1) the unperturbed dipole field (Campbell 1992; Livio & Pringle 1992)

$$B_z = B_* \left( \frac{R}{R_*} \right)^{-3}, \qquad (2)$$

where $R_*$ is the stellar radius and $B_* R_*^3 = \mu$, the stellar dipole moment. From this poloidal field, a toroidal component $B_\phi$ is generated from the shearing of field lines anchored both to the star and to the disc. The growth in $B_\phi$ due to shearing may be limited either by processes in the disc, for example by magnetic buoyancy or turbulent diffusion (Campbell 1992; Romanova, Lovelace & Bisnovatyi-Kogan 1995), or by reconnection of field lines in the magnetosphere. In the former case, the resultant value of the ratio $B_\phi/B_z$ is typically $\gg 1$, implying a strong twist of the magnetospheric field lines into what we consider to be an unstable configuration. Accordingly we assume reconnection in the magnetosphere to be the limiting mechanism, with a characteristic time-scale $\sim 1/\Omega$, where $\Omega$ is the Keplerian angular velocity of the disc material. The equilibrium value of the azimuthal field at the disc plane is then (Livio & Pringle 1992)

$$B_\phi = B_z [\Omega(R) - \Omega_*]/\Omega(R), \qquad (3)$$

vanishing at the corotation radius $R_\Omega$ and with the desired physical property that $B_\phi \sim B_z$ elsewhere. A radial field component $B_R$ can also be generated from $B_z$ by the inward dragging of field lines by the disc flow. However, as $v_R \ll v_\phi$ this component will be small provided that the magnetic diffusivity in the disc is of similar magnitude to the viscosity, as is likely to be the case (van Ballegooijen 1989; Lubow, Papaloizou & Pringle 1994). We assume that there is no significant $B_R$ component in our models.

### 2.2  Equation for disc surface density evolution

Equations (1) and (3) yield the magnetic torque experienced by an annulus at radius $R$. Incorporation of this into the standard derivation for the equation describing the time dependence of the surface density $\Sigma(R, t)$ in a Keplerian disc around a star of mass $M_*$ gives (e.g. Pringle 1981; Livio & Pringle 1992)

$$\frac{\partial \Sigma}{\partial t} = \frac{3}{R} \frac{\partial}{\partial R} \left[ R^{1/2} \frac{\partial}{\partial R} (\nu \Sigma R^{1/2}) \right] + \frac{1}{R} \frac{\partial}{\partial R} \left( \frac{\Omega - \Omega_*}{\Omega} \frac{B_z^2 R^{5/2}}{\pi \sqrt{GM_*}} \right), \qquad (4)$$

where $B_z$ is given by (2) and $\nu$ is the normal kinematic viscosity. The effect of the magnetic field is to add a second, advective term to the evolution equation, representing a magnetic torque which falls off very rapidly with increasing radius, and which causes inflow inward of corotation. Beyond corotation the field acts to inject angular momentum into the disc material. A sufficiently strong field can therefore cease inflow and accretion on to the star altogether.

In order to integrate equation (4), the form and magnitude of the kinematic viscosity $\nu$ need to be specified, the usual choice being the ad hoc $\alpha$ prescription relating the viscosity to the local disc sound speed $c_s$ and scale height $H$ via $\nu = \alpha c_s H$, with $\alpha$ a poorly known parameter (Shakura & Sunyaev 1973). With such an assumption it is possible to solve for the vertical thermal equilibrium structure of the disc, and hence to derive the viscosity as a function of the disc variables. Two recent independent calculations of this type for T Tauri discs by Kawazoe & Mineshige (1993) and Bell & Lin (1994) are in reasonable agreement. In common with earlier authors (Campbell 1992), we find that the disc is disrupted just inward of the radius where the magnetic torque exceeds the viscous torque. Thus at the inner edge the magnetic pressure is small compared with the thermal pressure (Clarke et al. 1995), and it is acceptable to use the results of the non-magnetic structure calculations even in the magnetic regime. For input mass fluxes in the region we are interested in ($\dot{M} \sim 10^{-7}$ $M_\odot$ yr$^{-1}$), the equilibrium curves given in Bell & Lin (1994) are well fitted by a single power law in $\alpha$, $R$ and $\Sigma$, corresponding to a viscosity given approximately by

$$\nu = 0.3 \alpha^{1.05} \Sigma^{0.3} R^{1.25}, \qquad (5)$$

which we use in equation (4). The value of $\alpha$ is highly uncertain, though arguments based on the recurrence timescales of FU Orionis outbursts (assuming these to be due to thermal instabilities in T Tauri discs) suggest a low value (Clarke, Lin & Pringle 1990; Kawazoe & Mineshige 1993; Bell & Lin 1994). For our 'best guess' runs we therefore take $\alpha = 10^{-3}$; we also consider $\alpha = 10^{-1}$, this higher value being typical of that deduced in dwarf nova systems (e.g. Cannizzo 1993 and references therein).

### 2.3 Numerical method

Equation (4) is integrated on an Eulerian radial grid evenly spaced in $R^{1/2}$, using a simple first-order explicit scheme for the diffusive term and the Lelevier method (Potter 1973) for the magnetic torque term. With the outer boundary of the disc at 120 $R_\odot \approx 0.6$ au, a few hundred grid points suffice to resolve adequately the comparatively narrow region over which the magnetic torque has a significant influence. If the disc is disrupted by the field, we cut off the calculation when the advective inflow velocity exceeds some limiting value $v_{\rm max}$, thereby preventing the time-step becoming arbitrarily small at the disc inner edge. We find that the value of the parameter $v_{\rm max}$ is unimportant provided that it greatly exceeds the typical viscous inflow velocity, so that the magnetic torque dominates the viscous torque well before $v_R > v_{\rm max}$. The calculations described here use $v_{\rm max} = 5 \times 10^5$ cm s$^{-1}$, so that the calculation is cut off at a radius where the inflow velocity is first becoming roughly supersonic. We define this radius as the magnetospheric radius $R_{\rm m}$. Although by this radius the assumptions used to derive equation (4) are breaking down, the steep radial dependence of the magnetic torque implies that a more detailed treatment of the inner disc region would not change $R_{\rm m}$ substantially.

At the outer boundary we impose a given mass flux $\dot{M}_{\rm outer}$. At the inner boundary we allow flow through the inner edge if $R_{\rm m} \leq R_\Omega$. If $R_{\rm m} > R_\Omega$, the inner boundary condition must be set so that there is no accretion on to the star. Integration of the magnetic part of equation (4) with respect to $R$ gives

$$\int_{R_{\rm in}}^{R_{\rm out}} \frac{\partial}{\partial t} (2\pi R \Sigma) \, {\rm d}R = 2\pi [f(R)]_{R_{\rm in}}^{R_{\rm out}}, \qquad (6)$$

where

$$f(R) = \frac{\Omega - \Omega_*}{\Omega} \frac{B_z^2 R^{5/2}}{\pi \sqrt{GM_*}}. \qquad (7)$$

The left-hand side of equation (6) is $\partial M_{\rm disc}/\partial t$, which for $R_{\rm m} > R_\Omega$ and large $R_{\rm out}$ must be zero when only the magnetic term is considered. This ensures that the disc mass in the computational region is conserved (note that the diffusive part is solved using a mass-conserving scheme and so does not alter this condition). The steep radial dependence of $B_z$ means that $f(R)|_{R_{\rm out}} = 0$, and hence the correct boundary condition is obtained by setting $f(R) = 0$ at the inner edge as well. As a consequence of cutting off the calculation, the innermost zone (where $v_R$ may exceed $v_{\rm max}$) will not generally be computed correctly, and the calculated surface density may become negative. When this occurs we restore the surface density to zero, and compensate by removing mass from the first positive zone.

The stellar mass, radius and spin rate will vary with time as a consequence of both accretion and the coupling between the star and its disc. Even without such interaction, normal pre-main-sequence evolution will lead to changes in the stellar parameters. In a recent paper Clarke et al. (1995) estimated that, for a T Tauri star accreting at $\sim 10^{-7}$ M$_\odot$ yr$^{-1}$, the time-scale for changes in the spin period was $\sim 10^5$ yr; this is evidently shorter than the time-scales for changes in either $M_*$ or $R_*$. As we are interested here in magnetic field variations on time-scales $\ll 10^5$ yr, it suffices for our purposes to regard the stellar parameters as fixed for the duration of the calculation.

As the inner regions of T Tauri discs are extremely optically thick, we compute the disc emission by assuming that a blackbody spectrum is radiated at each annulus, with an effective temperature $T_{\rm eff}$ found by equating the rate of radiative energy loss to the energy generation at each radius. We consider two limits, first that where the work done by the magnetic torque is dissipated in the magnetosphere and does *not* go into heating the disc. In this limit the disc luminosity is provided entirely by kinematic viscosity, giving

$$2\sigma T_{\rm eff}^4 = \frac{9}{4} \nu \Sigma \Omega^2, \qquad (8)$$

where $\sigma$ is the Stefan-Boltzmann constant. In the alternative limit, where all the work done by the field is returned to reheat the disc material, equation (8) becomes

$$2\sigma T_{\rm eff}^4 = \frac{9}{4} \nu \Sigma \Omega^2 + \frac{|B_z B_\phi|}{2\pi} R \Omega, \qquad (9)$$

where $B_z$ and $B_\phi$ are given by equations (2) and (3). Here $\nu(\Sigma)$ does not include the effect that the magnetic heating may have on the kinematic viscosity. Outside a narrow inner area of the disc (over which the magnetic torque is anyway dominant) such an effect should be small. From the run of $T_{\rm eff}$ with radius we calculate in each limit the bolometric disc luminosity $L_{\rm bol}$ and the fluxes at 2.2 and 5 $\mu$m, these wavelengths being characteristic of the temperatures at the inner edge of the disc.

We note here that, as Keplerian rotation is assumed throughout, even our 'reheating' limit does not include the energy dissipation that must occur to brake the disc material to $\Omega_*$ and permit it to free-fall on to the star. Such braking presumably occurs just inward of our calculated $R_{\rm m}$, where the density is very low, and thus the resulting emission is likely to be optically thin. Provided that this is the case, the energy will emerge at the higher frequencies and in the emission lines where the accretion shock component is dominant (compared to which the braking energy is negligible for large $R_{\rm m}$). Hence little error is incurred in ignoring it when calculating the disc luminosity.

## 3 NUMERICAL RESULTS

For typical T Tauri stellar parameters we take mass $M_* = 1$ M$_\odot$, radius $R_* = 3$ R$_\odot$, with an input mass flux into the outer edge of the disc $\dot{M}_{\rm outer} = 10^{-7}$ M$_\odot$ yr$^{-1}$. Stellar rotation periods are in the range 4–12 d, similar to the range seen in photometric surveys of CTTs (Bouvier et al. 1993; Edwards et al. 1993). As there is no real observational or theoretical guidance for the time dependence of the field, we adopt the simple form

$$B_* = B_{\rm dc} + B_{\rm ac} \sin\left(\frac{2\pi t}{\tau_{\rm cycle}}\right), \qquad (10)$$

envisaging both a 'frozen-in' component $B_{\rm dc}$ with no time dependence and an oscillatory dynamo component $B_{\rm ac}$, with period $\tau_{\rm cycle}$. The cycle times of interest observationally will have $\tau_{\rm cycle} \sim 10$ yr; activity on this time-scale appears to be common in stars surveyed as part of the Mt Wilson monitoring programme (Baliunas 1988).

The behaviour of the star-disc system in this model divides into two qualitatively different regimes. We first consider a variable field such that the magnetosphere lies within the corotation radius throughout the cycle. In this case, although the magnitude of accretion on to the star may be strongly modulated, some accretion does occur at all phases of the cycle. For $\dot{M}_{\rm outer} = 10^{-7}$ M$_\odot$ yr$^{-1}$ and rotation period of 6 d, we find that magnetic fields between a few hundred gauss and a kilogauss will disrupt the disc yet leave $R_{\rm m}$ always within corotation. Such a situation should be favoured in the long-term evolution of the system as it can correspond to a state where the time average of the torque on the star is zero. Calculations of the spin evolution of magnetic T Tauri stars (Cameron & Campbell 1993; Cameron et al. 1995) suggest that for much of the T Tauri phase the stellar spin period can adjust to this equilibrium state.

The second regime we examine is that where the magnetospheric radius crosses corotation during the course of the cycle. In this case accretion ceases altogether when $R_{\rm m} > R_\Omega$, and matter can flow on to the star only in pulses during the weak-field phases of the cycle. This requires a stronger average field of $\sim 2$ kG. This cannot be an equilibrium state, as the magnetic field will exert an average spin-down torque that reduces the stellar rotation rate until the magnetosphere lies within corotation. However, such out-of-equilibrium states do occur in the spin evolution models referred to above. Moreover, the erratic behaviour of the solar activity cycle suggests that it is not unreasonable to envisage periods when the activity is much stronger or weaker than the long-term average.

### 3.1 Magnetic cycles inside corotation

Fig. 1 shows results from a typical calculation where the magnetosphere remains within corotation throughout the cycle. Here $B_{\rm dc} = 600$ G, $B_{\rm ac} = 300$ G, $\tau_{\rm cycle} = 10$ yr and the stellar period was 12 d. The calculation was run until a periodic state was reached and the effects of the (arbitrary) initial conditions had been erased. For the $\alpha = 10^{-3}$ run shown, this required a few times $10^4$ years. We plot the accretion rate through the inner edge of the disc in response to the applied stellar field, and the bolometric luminosity of the disc in the two limits discussed above.

The results show a strong modulation of the accretion rate on to the star, with the ratio $\dot{M}_{\rm max}/\dot{M}_{\rm min}$ exceeding 8. This is in phase with the applied stellar magnetic field. For the disc flux, the degree of variation predicted depends on the assumption made as to where the magnetic work is deposited. If this is primarily in the magnetosphere (the 'no reheating' limit) then no significant variation is expected. Conversely, if all the energy is deposited eventually into the disc then modulation of the bolometric disc luminosity by a factor of $\sim 1.5$ is seen, corresponding to a change in the disc emission at 2.2 $\mu$m of $\sim 3$ mag. The 5-$\mu$m variation, sampling material further out that is less affected by the cycle, is 0.6 mag in this limit.

The negligible variation in disc flux in the no-reheating limit is a consequence of the fact that $R_{\rm m}$ is nearly constant during the cycle. This is because the amount of mass accreted during the strong-field phase of the cycle is small compared to the disc mass, so that the inner edge is unable to sweep back and forth by more than a small amount. For an estimate, we ignore viscosity (which will reduce variations in $R_{\rm m}$) and note that in a periodic state the mass accreted in one magnetic field cycle is $\dot{M}_{\rm outer}\tau_{\rm cycle}$. Comparison of this with the mass swept up in displacing $R_{\rm m}$ outward by $\Delta R_{\rm m}$, $2\pi R_{\rm m}\Delta R_{\rm m}\Sigma$, where $\Sigma$ is some characteristic surface density near the inner edge, gives

$$\frac{\Delta R_{\rm m}}{R_{\rm m}} \sim \frac{\dot{M}\tau_{\rm cycle}}{2\pi R_{\rm m}^2 \Sigma}. \quad (11)$$

As the surface density in the disc scales approximately as $\alpha^{-1}$, low values of $\alpha$ lead to high surface density and small radial excursions in $R_{\rm m}$. For an $\alpha$ of $10^{-3}$ the surface density reaches $\sim 7 \times 10^3$ gcm$^{-2}$ within 2 R$_\odot$ of the disc inner edge, and as a consequence $\Delta R_{\rm m}/R_{\rm m} \ll 1$ for $\tau_{\rm cycle} = 10$ yr. The surface density profile and viscous dissipation then remain essentially fixed throughout the cycle, and little modulation of the disc luminosity occurs unless the variable magnetic work term is included.

From equation (11) it is apparent that the cycle time is an important parameter in determining the behaviour. To investigate this, a variety of models were computed with $\tau_{\rm cycle}$ between 10 and $10^4$ yr. Long cycle periods are expected to reduce the modulation of the accretion rate, as for a very slow cycle the disc has time to adjust via viscous processes to the changing field and remains close to a steady state with $\dot{M}_{\rm inner} \approx \dot{M}_{\rm outer}$. Fig. 2 shows the results of varying $\tau_{\rm cycle}$, keeping all other parameters fixed as before, for $\alpha = 10^{-3}$. The plot shows the magnitude of variation in the accretion rate $f = \dot{M}_{\rm max}/\dot{M}_{\rm min}$, and in the disc bolometric luminosity for either limit. As anticipated, the modulation of accretion drops off sharply as the cycle time approaches the viscous response time of the disc (here $\sim 10^4$ yr at the outside edge). For this low value of $\alpha$ it is apparent that all time-scales of observational interest are 'fast' compared to the disc time-scale, and will thus yield substantial modulation similar to that described above.

For a fast cycle it is straightforward to estimate how the variability in the accretion rate scales with the magnitude of the oscillatory component in the field. In the limit of a very rapid cycle, $R_{\rm m} \to$ constant, and the disc surface density profile remains essentially fixed over the cycle. Near $R_{\rm m}$ the magnetic field dominates viscosity in transporting angular momentum, hence the accretion rate scales as the magnetic torque which from equations (1) and (3) is proportional to $B_*^2$. We then have

$$\frac{\dot{M}_{\rm max}}{\dot{M}_{\rm min}} = \left(\frac{B_{\rm max}}{B_{\rm min}}\right)^2. \quad (12)$$

For a low value of $\alpha$ and $\tau_{\rm cycle} \sim 10$ yr we find that this simple relation agrees reasonably with the computed results, though as there is always *some* movement of $R_{\rm m}$ the calculated variation is slightly less than this extreme case. We note that substantial modulation of the accretion on to the star occurs even for $B_{\rm ac}$ considerably smaller than $B_{\rm dc}$.

An increase in the value of $\alpha$ decreases the disc mass and response time-scale, and therefore decreases the vari-ability in $\dot{M}_{\rm inner}$ for any given magnetic cycle period. We find that an increase in the value of $\alpha$ to $10^{-1}$ while leaving all other parameters unaltered reduces the magnitude of variation in accretion rate by approximately a factor of two. The qualitative behaviour is identical to the low-$\alpha$ calculation. In this case there is substantial movement of $R_{\rm m}$, and hence variation occurs in the disc luminosity for both the no-reheating and full-reheating assumptions. Depending on which assumption is made, we find variation in the disc bolometric luminosity to be between $\sim 5$ per cent and $\sim 30$ per cent. The upper limit is about 2 mag at 2.2 $\mu$m, and 0.5 mag at 5 $\mu$m.

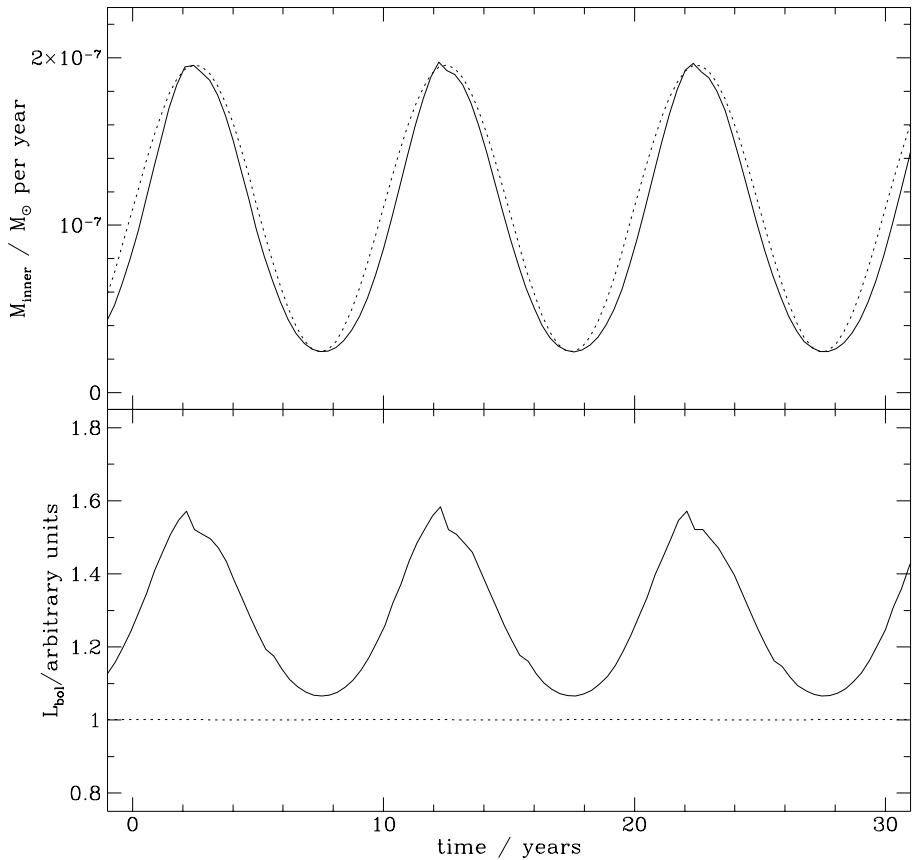

**Figure 1.** Modulation of accretion rate and disc luminosity for cycle with $R_{\rm m} < R_\Omega$ throughout. Top panel shows accretion rate (solid line) with the applied stellar field (dotted line) to illustrate phase. Bottom panel shows bolometric disc flux for the reheating (solid line) and no-reheating (dotted line) limits.

### 3.2 Magnetic cycles crossing corotation

An example of the behaviour when the magnetospheric radius crosses corotation during a cycle is shown in Fig. 3. For this run $B_{\rm dc} = 3$ kG and $B_{\rm ac} = 1.5$ kG; with a stellar spin period of 4 d this easily disrupts the disc out to corotation. Again the calculation was run until a periodic state was reached. We show the accretion rate, the applied magnetic field and the bolometric luminosity of the disc. Here $\tau_{\rm cycle} = 10$ yr and $\alpha$ was taken as $10^{-1}$. Accretion occurs in a series of relatively narrow pulses during the weak-field phases of the cycle when the magnetosphere passes within the corotation radius. The fraction of the cycle for which accretion occurs is $\sim 0.4$, with a peak accretion rate $\dot{M}_{\rm inner} \sim 4 \times 10^{-7}$ M$_\odot$ yr$^{-1}$, i.e. four times the steady input flux at the outer edge. The corresponding variation in $L_{\rm bol}$ is $\sim 10$ per cent if energy is deposited in the magnetosphere, but this rises to approximately a factor of two if that energy is used to reheat the disc.

The equivalent calculation for $\alpha = 10^{-3}$ yields very similar results. In this case we find a peak accretion rate of $\dot{M}_{\rm inner} \sim 3 \times 10^{-7}$ M$_\odot$ yr$^{-1}$, with a duty cycle of $\sim 0.6$. In general we find that, for cycles crossing corotation, the details of the cycle and the value assumed for $\alpha$ are much less important than when the magnetosphere remains within corotation. This is because, for cycles crossing the corotation radius, the qualitative behaviour is determined by the switching on and off of accretion as $R_{\rm m}$ passes $R_\Omega$. Hence we find similar variations in accretion rate both for lower values of $\alpha$ and for cycles with smaller magnetic field variation $B_{\rm ac} = 750$ G, although the latter value does reduce the likely modulation of the disc luminosity.

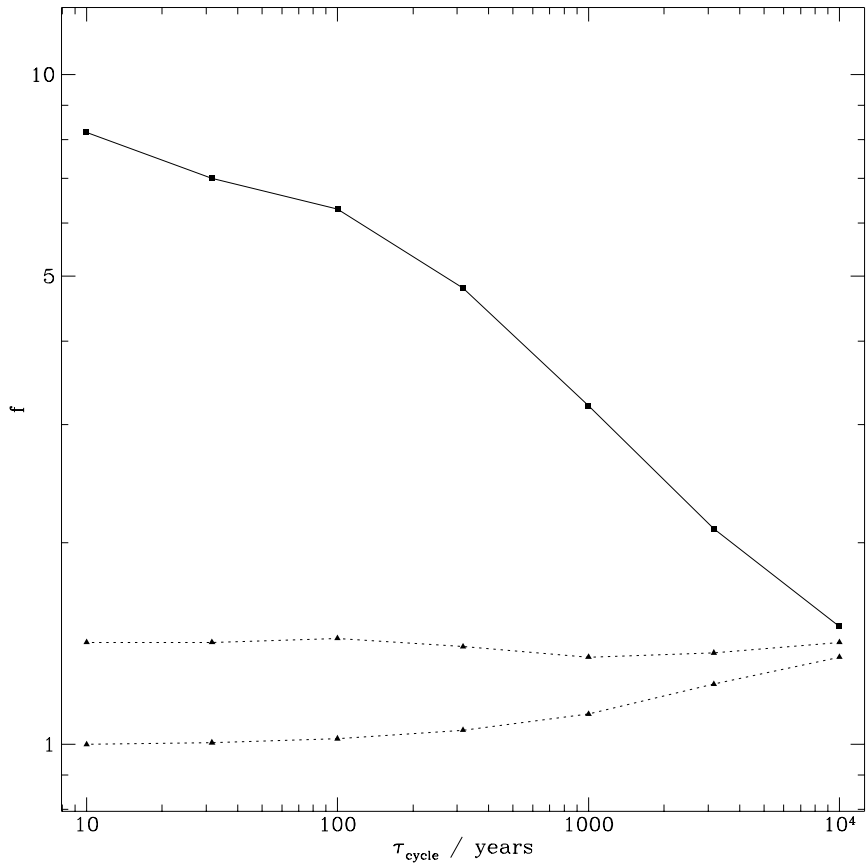

**Figure 2.** Variation of ratio of maximum to minimum flux $f$ as a function of magnetic cycle period. Solid line shows modulation of the accretion rate; dotted lines bracket the variation in disc luminosity obtained from the limiting cases.

## 4   DISCUSSION

We have shown that cyclic stellar magnetic activity can lead to various patterns of variability for the star-disc system. The most robust prediction is that cycles with periods of years to decades should strongly modulate the rate of accretion on to the central star. We find this to be the case regardless of the location of the magnetosphere relative to corotation, and for all values of the $\alpha$ parameter considered plausible in T Tauri discs. Observationally this translates into variability in the UV radiation emitted in the accretion shock, and in the spectral lines believed to be formed in the infalling accretion column. Our model is unable to predict the details of this emission, which for a real system will depend on the structure of the field close to the photosphere and the inclination of the magnetic and spin axes. Modelling of photometric data (Bouvier et al. 1993; Kenyon et al. 1994) suggests hotspots with effective temperatures up to $\sim 10^4$ K covering from a few to 30 per cent of the stellar surface. With such temperatures the large changes in $\dot{M}_{inner}$ found above would probably lead to variations in $UBV$ fluxes of several magnitudes over the course of a cycle. The strong emission lines in this spectral range render any more definite prediction of the variability questionable.

For the particular case where the magnetosphere crosses the corotation radius in the course of a cycle, the accretion on to the central star occurs in pulses, corresponding to the weak-field phases of the cycle. Such an accretion event is characterized by a rapid rise or fall in the apparent accretion rate, with a peak rate that can be several times larger than that found by averaging over the cycle. During the period when accretion is switched off, the star would appear to be 'weak' in terms of indicators of accretion close to the photosphere, but would still possess a reservoir of material held up by the field at a few stellar radii (Clarke et al. 1995). Such a system could revert to 'classical' status when the field weakened.

The variability of the disc flux during the cycle depends both on the value of $\alpha$ and on the (unknown) details of the interaction between the disc and the magnetosphere. Accordingly we can only place limits on the magnitude of variability in the near-infrared wavelengths emitted by the inner disc. Significant variations of perhaps $1 - 2$ mag at 2.2 $\mu$m seem probable provided that a large fraction of the work done by the magnetic field goes into heating the disc material. Because the field affects only the inner part of the disc, any variability in disc flux due to a magnetic cycle drops away rapidly at longer wavelengths.

A stellar magnetic field that varies by order unity may cause further variability in the star-disc system by affecting the photospheric output of the star. During the strong-field phases, convection may be inhibited and the star should dim (Appenzeller & Dearborn 1984). Variations similar to those discussed here are estimated to produce changes in visual brightness of a few magnitudes. Such an effect would be most significant in systems with relatively low accretion rates, where the photospheric component is consequently strong.

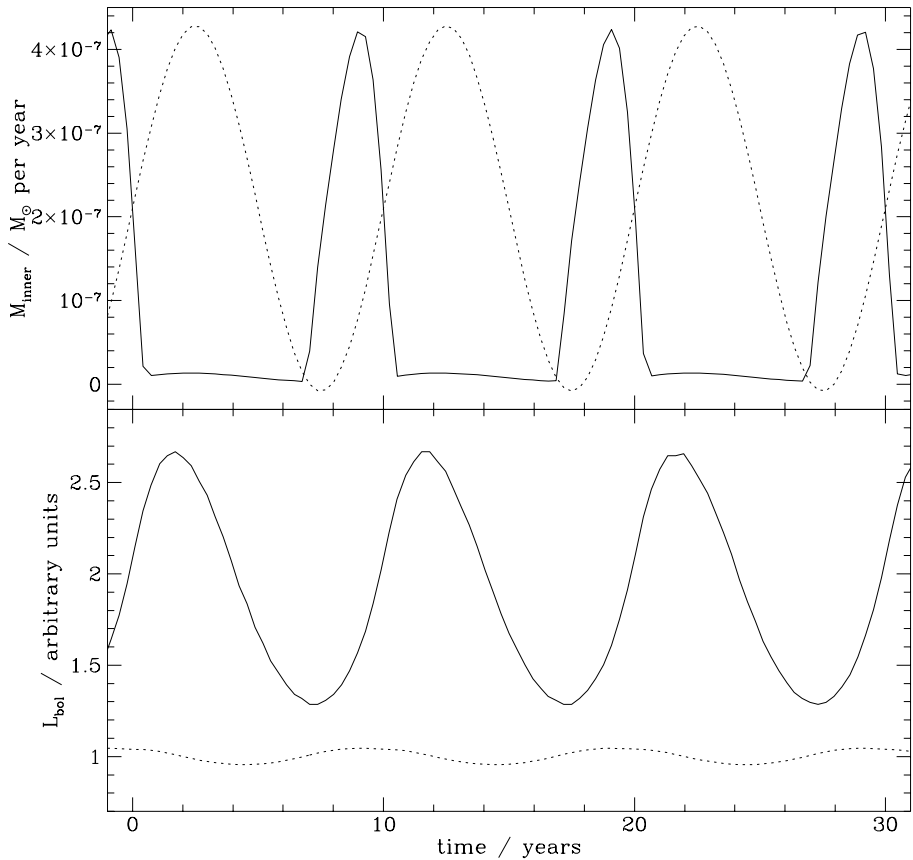

**Figure 3.** Modulation of accretion rate and disc flux for a cycle crossing the corotation radius. Top panel shows accretion rate (solid line) with the applied stellar field (dotted line) to illustrate phase. Bottom panel shows bolometric disc flux for the reheating (solid line) and no-reheating (dotted line) limits.

Here we have considered strictly periodic stellar magnetic behaviour, whereas analogy with the Sun might suggest a much richer array of quasiperiodic changes. We therefore note that the cause of the large modulation of accretion seen in our calculations is the mismatch between the magnetic cycle time-scale and that of the viscous response of the disc, and is thus independent of the details of the magnetic variation. Hence, in any model where the disc is disrupted by a stellar field, changes in that field that occur on observationally accessible time-scales can be expected to lead to variations in the accretion rate through the inner edge of the disc.

As a result of the large magnitude of variability discussed here, we conclude that useful constraints could be placed on the amplitude of magnetic cycles from observations of pre-main-sequence stars, even given the inevitable 'noise' caused by the short-time-scale variations in these objects. The large archival dataset available on T Tauri stars may then constitute an indirect record of such behaviour stretching back many decades. Detection of such cycles would both provide an explanation for decade-scale variability in T Tauri stars, and extend greatly the range of stellar systems in which magnetic dynamo activity may be studied.

## ACKNOWLEDGMENTS

I thank Cathie Clarke and Jim Pringle for their assistance with all aspects of this work.